# Drive Right: Autonomous Vehicle Education Through an Integrated Simulation Platform

Zhijie Qiao, Helen Loeb, Venkata Gurrla, Matt Lebermann, Johannes Betz, Rahul Mangharam

School of Engineering and Applied Science, University of Pennsylvania

Email: zhijie@seas.upenn.edu

**Abstract**

Autonomous vehicles are being rapidly introduced into our lives. However, public misunderstanding and mistrust have become prominent issues hindering the acceptance of these driverless technologies. The primary objective of this study is to evaluate the effectiveness of a driving simulator to help the public gain an understanding of autonomous vehicles and build trust in them. To achieve this aim, we built an integrated simulation platform, designed various driving scenarios, and recruited 28 participants for the experiment. The study results indicate that a driving simulator effectively decreases the participants' perceived risk of autonomous vehicles and increases perceived usefulness. The proposed methodologies and findings of this study can be further explored by auto manufacturers and policy makers to provide user-friendly autonomous vehicle design.

## I. Introduction

While billions of dollars have been invested in autonomous driving (AD) technology, little work has been done to prepare the public for this paradigm shift. As auto manufacturers press forward with the introduction of ever more advanced driver assistance features, the concept of AD still sounds unsettling to many people. According to a survey of 1,200 adult drivers conducted by the Partners for Automated Vehicle Education (PAVE), 48 % Americans said they "would never get into a taxi or ride-share vehicle that was driven autonomously", while 20 % believed autonomous vehicles (AVs) would never be safe [1]. Another survey conducted by AAA of over 1,000 U.S. adults found that 54 % of all participants were afraid to ride in an autonomous vehicle, while 32 % were unsure about it [2]. AVs have the potential of saving millions of lives from needless traffic accidents and could drastically reduce the cost associated with the transportation industry. However, public mistrust and the drivers' reluctance to relinquish control have become prominent issues hindering the acceptance of these driverless technologies [3].

Researchers from MIT AgeLab found that most people, including the elder generation, were comfortable with the idea of technological innovation in the driving industry. However, improved training methods and preferred training strategies played an important role in the eventual adoption of the technology [4]. Further, drivers must receive information that helps them explain and predict the vehicle's behavior [5]. To properly demonstrate AVs, one would need to sit in an actual car, with an experienced human supervisor monitoring the safety and explaining the vehicle's operation. However, such demonstrations can be expensive and take a considerably long time. Moreover, many people may not feel comfortable stepping into an autonomous vehicle before they fully trust its performance [6]. In such circumstances, a driving simulator provides an alternative approach to demonstrate AVs in a safe and controllable environment. Driving simulators are also time-efficient, cost-effective, and can be used in a variety of places.

In this study, we developed an autonomous vehicle simulation and demonstration system. Our system leveraged the latest simulation tools and autonomous driving platforms: SVL Simulator [7, 8], Baidu Apollo [9], and Autoware Auto [10, 11]. This integrated system was designed to improve the public's understanding of AD technology and help them build trust. The paper is structured as follows: Section II provides a literature review of the existing efforts made by the academic community to explore various user-AV interactions. Section III describes our simulator system design and the test scenario development. Section IV presents the procedures of our human study. Section V presents the results, and Section VI discusses the implications. Finally, Section VII summarizes the findings and provides insights for future development.

## II. Literature Review

*Information Delivery:* extensive research has been performed to evaluate the kind of AD information that should be provided to passengers and in what form it should be provided. Koo et al., in their human machine interaction research, discovered that the "why" information explaining the vehicle's behavior is more important than the "how" information, reaffirming the vehicle's action [12]. Morra et al. found that the AD system should display a complete picture of the vehicle's surrounding environment including other vehicles, pedestrians, and traffic indicators. This informative approach, although more cognitively demanding, could contribute to a less stressful riding experience [13]. Similar results have been found in Haeuslschmid's AD trust research. In their study comparing three visualization methods, "A world in miniature" was preferred by most people and received the highest score of trust. According to the researchers, this method presented the "car's perception of the surroundings, its interpretation, and its actions in a clear and competent way" [14].

Besides providing reasoning to justify the operation of autonomous vehicles, researchers have conducted experiments to exploit the mental and psychological benefits of AV system design. For instance, Sun and Zhang proposed their synesthetic-based multimodal interaction (SMBI) model, which utilized voice and lightning to raise the driver's awareness. Under emergent conditions, the vehicle's speech prompt changes from low to high frequency, and the ambient light changes from blue to red. Sun and Zhang found the drivers were more likely to hold the steering wheel and pay attention to the road when experiencing this sudden ambient change. Drivers also reported higher scores towards this interactive system design in terms of trust, technical competence, situational management, and perceived ease of use [15].

*Anthropomorphism:* another important aspect of the AV human interaction is the concept of anthropomorphism. In an anthropomorphic design, the vehicle is imbued with humanlike characteristics, motivations, intentions, or emotions [16]. Providing human-like features is a common approach to increase trust and acceptance in non-human agents [17]. Research has shown that adding a humanized conversational interface alongside the traditional graphics interface could increase system transparency and portray the vehicle as "smart" [18]. It is also suggested that representing the vehicle's symbolic indicators (e.g., right turn, go straight, accelerate) as animated facial movements would increase user liking and trust of the system [19]. All these experimental results indicate that anthropomorphic design could be an effective approach to help users build trust and confidence in AVs.

*Simulated Riding Experience*: the most efficient way for users to gain understanding of AVs is to ride in an actual vehicle and interact with it. However, these demonstrations are expensive and time-consuming. Moreover, the real-road test poses potential physical and psychological harm to the participants. In response to this challenge, various driving simulators have been introduced to test and demonstrate AVs in a safe and controllable environment. For instance, Dosovitskiy et al. designed an open urban driving simulator called CARLA that aimed to support the "development, training, and validation" of AVs [20]. Best and his colleagues built an AD simulation platform "AutonoVi-Sim" that focused on weather, sensing, and traffic control [21]. Manawadu and his team used a simulator to study the driving performance of humans versus AVs. They found that on average, the autonomous vehicle decreased the task completion time, rate of collision, and driver's mental workload [22].

Building an AD system that follows conventional traffic norms and driving styles has also proven to be important. Drivers and passengers expect AVs to make intelligent decisions and act like human agents. For instance, a study showed that users reported higher perceived risk when an autonomous vehicle drove slowly on a clear day and lower perceived risk when the same vehicle drove slowly on a snowy night [23]. Furthermore, Sun et al. demonstrated that building a personalized vehicle that mimics the driver's behavior could reduce the perceived risk and increase perceived usefulness. In their study, the drivers' driving data was recorded to build a personalized AD system. In the subsequent experiments, the personalized AV received higher scores in terms of trust, comfort, and situational awareness compared to the standardized AV [24].

A clear limitation of the driving simulator is the users' awareness of the simulation environment and their potential bias during the engagement. To create a real-road autonomous driving experience, researchers at Stanford University introduced their RRADS Platform: "A Real Road Autonomous Driving Simulator". In their setup, a driving wizard (human driver hidden from passengers) mimicked the control of an autonomous vehicle, and an interaction wizard (researcher that assisted the participants) explained the vehicle's operation. While this innovative design provided some real-road AD experience, the wizard's driving style clearly affected the passengers' attitude towards AVs [25].

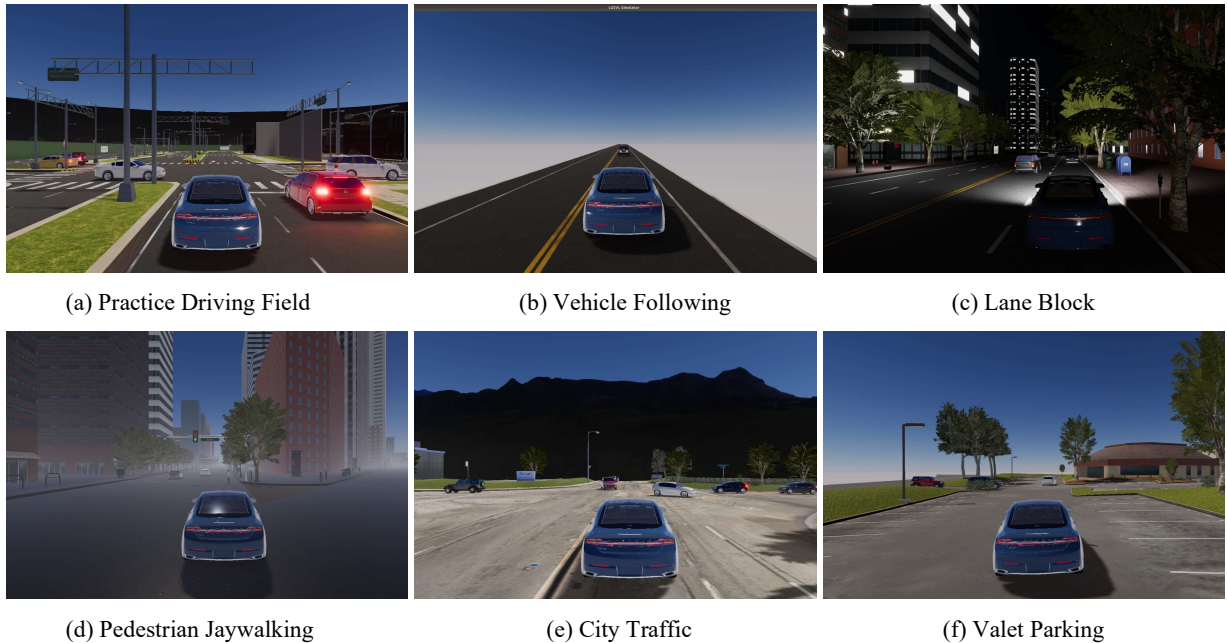

(a) Practice Driving Field  (b) Vehicle Following  (c) Lane Block
(d) Pedestrian Jaywalking  (e) City Traffic  (f) Valet Parking

**Figure 1.** Practice driving field and five testing scenarios of the SVL Simulator

## III. Methods

*SVL Simulator:* the SVL Simulator is an open-source driving simulator designed to facilitate the development of AD research. Built with the game engine Unity, the simulator allows the construction of 3D digital twins of the real world using point cloud and image data. The simulator supports multiple real-time sensor inputs and outputs including the Camera, LiDAR, Radar, GPS, and IMU. Environmental parameters can also be adjusted, such as time of the day, weather, vehicles and pedestrians using a Python API interface. With internal bridge support such as the Robotic Operating System (ROS), ROS2, and CyberRT, the simulator can be connected to popular AD platforms like Apollo and Autoware.

*Logitech G920:* Logitech G920 is a driving force steering wheel and pedal set. Its full throttle, full control capability makes it suitable for various driving tasks. In our simulation, this steering wheel pedal set is connected to the SVL Simulator to provide a high fidelity driving experience.

*Apollo 5.0:* Apollo is an open-source AD platform developed by the leading technology company Baidu. It is an industrially used Level 4 AD platform as defined by the Society of Automotive Engineers (SAE) [26]. Apollo's MinuBus project launched massive production in 2018, and its Robo Taxi project is the first attempt to use AD in the commercial transportation business. The latest Apollo 5.0 version (Apollo 6.0 is available, but still under modular testing) includes a set of integrated modules, such as a Map Engine, Localization, Perception, Prediction, Planning, and Control. These modules coordinate with each other to provide a safe and reliable driving experience. Apollo's CyberRT bridge allows for a connection to the SVL Simulator for AD testing and demonstration. Its web based Dreamview interface displays information in real time for user-friendly visualization and debugging.

*Autoware Auto*: Autoware is an open-source software stack introduced by the Autoware Foundation. Its latest version, Autoware Auto, aims to address the problem of valet parking and autonomous cargo delivery. Autoware uses LiDAR for vehicle localization and motion planning. Its hardware and software have been successfully integrated in a Lexus vehicle, which could then perform valet parking using the mobile application control. In our implementation, Autoware is connected to the SVL Simulator using the ROS2 bridge and controlled using the rviz2 graphics interface.

*Testing Scenarios:* for the research study, five testing scenarios (Figure 1) were developed in the SVL Simulator using different maps and environmental settings. The first four scenarios were successfully completed using Apollo, and the valet parking scenario was achieved using Autoware. The testing vehicle was a 2017 Lincoln MKZ. To run AD, two gaming laptops were used, both equipped with an Intel i7 Processor and an Nvidia GTX 1080 Graphics Card. The operating system was Ubuntu 20.04. One laptop was set to run the SVL Simulator, and the other AD platform. The two machines were connected using an ethernet cable for real-time communication. This hardware setup provided sufficient computing power to ensure that both the simulator and the AD platform run smoothly.

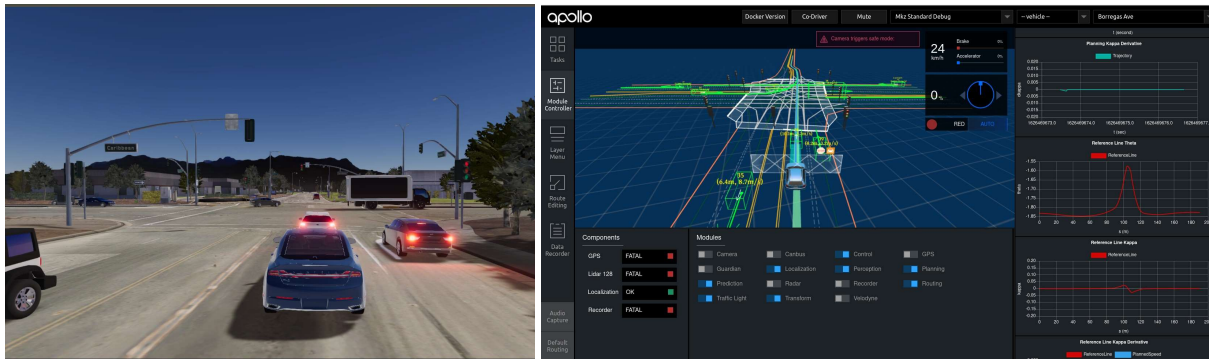

**Figure 2.** SVL Simulator (left) and Apollo Dreamview (right). Apollo Dreamview displays the ego vehicle's routing information, planning and control graph, sensing and prediction of moving objects, and responses to traffic indicators.

**Vehicle Following**: the ego vehicle needs to follow a reckless vehicle on a straight, single lane road. The reckless vehicle drives at a non-constant speed and occasionally makes sudden stops. The ego vehicle must respond quickly to the front vehicle's speed changes, in order to avoid a rear-end collision.

**Lane Block**: the ego vehicle drives on the right lane of a two-lane road, but is soon blocked by an illegally stopped vehicle. The ego vehicle must slow down first, yield to passing vehicles on the left lane, and then switch lanes to proceed driving. This scenario is simulated on a dark night, which makes it difficult to see the illegally stopped vehicle from afar.

**Pedestrian Jaywalking**: the ego vehicle must brake quickly to avoid a jaywalking pedestrian emerging from the side of the road. Time-to-collision is set at 2.6 seconds if no action is taken and the ego vehicle maintains its speed. This scenario is set on a rainy and foggy day, which increases the difficulty of seeing the pedestrian.

**City Traffic**: the ego vehicle needs to drive through an urban area with high density city traffic. While driving, the ego vehicle must follow all traffic rules and indicators. The driving route includes a four-way intersection with traffic lights and an unprotected left turn. The simulation map is a digital twin of a real street block in Sunnyvale, CA.

**Valet Parking**: the ego vehicle starts at the drop-off location. It needs to drive through the parking lot, reach the designated parking spot, and perform reverse parking. While driving, the ego vehicle must avoid hitting other vehicles and pedestrians. The simulation map is a digital twin of a real parking lot in San Jose, CA.

*Participants:* we recruited 28 participants via email and social media. The only requirement to enter the study was the possession of a valid driver's license. Our participant group consisted of 16 males, 12 females and had a mean age of 25.2 years. The participants were not compensated but informed that they could potentially enhance their understanding of AVs. This study was approved by the Institutional Review Board (IRB) and all participants gave their informed written consent.

In the human study, participants were asked to drive through the above scenarios manually using the steering wheel and pedal set. The scenarios were presented in the third-person view, as shown in the above figures. The steering wheel and pedals were set up with proper force feedback, mimicking those of a real vehicle, and their control inputs were processed quickly by the simulator without any visible delay. In addition, the participants could use buttons on the steering wheel to activate common features such as: the headlight, the taillight, the turn signal, or gear shift. After the participants completed each scenario, we demonstrated how AVs approached the same situation using Apollo or Autoware.

The primary objective of running this driving experiment is to help human drivers better evaluate the performance of AVs. By taking part in the experiment, participants were made aware of how challenging some scenarios can be for human drivers. By watching the autonomous vehicle smoothly handling these same exact situations, the participants gain an understanding of the value of AD technology. The assumption is that this increased awareness will translate into a gain of trust and confidence. Notably, we are not trying to demonstrate that AVs' performance is superior to that of human drivers, a fact that could only be established with a robust

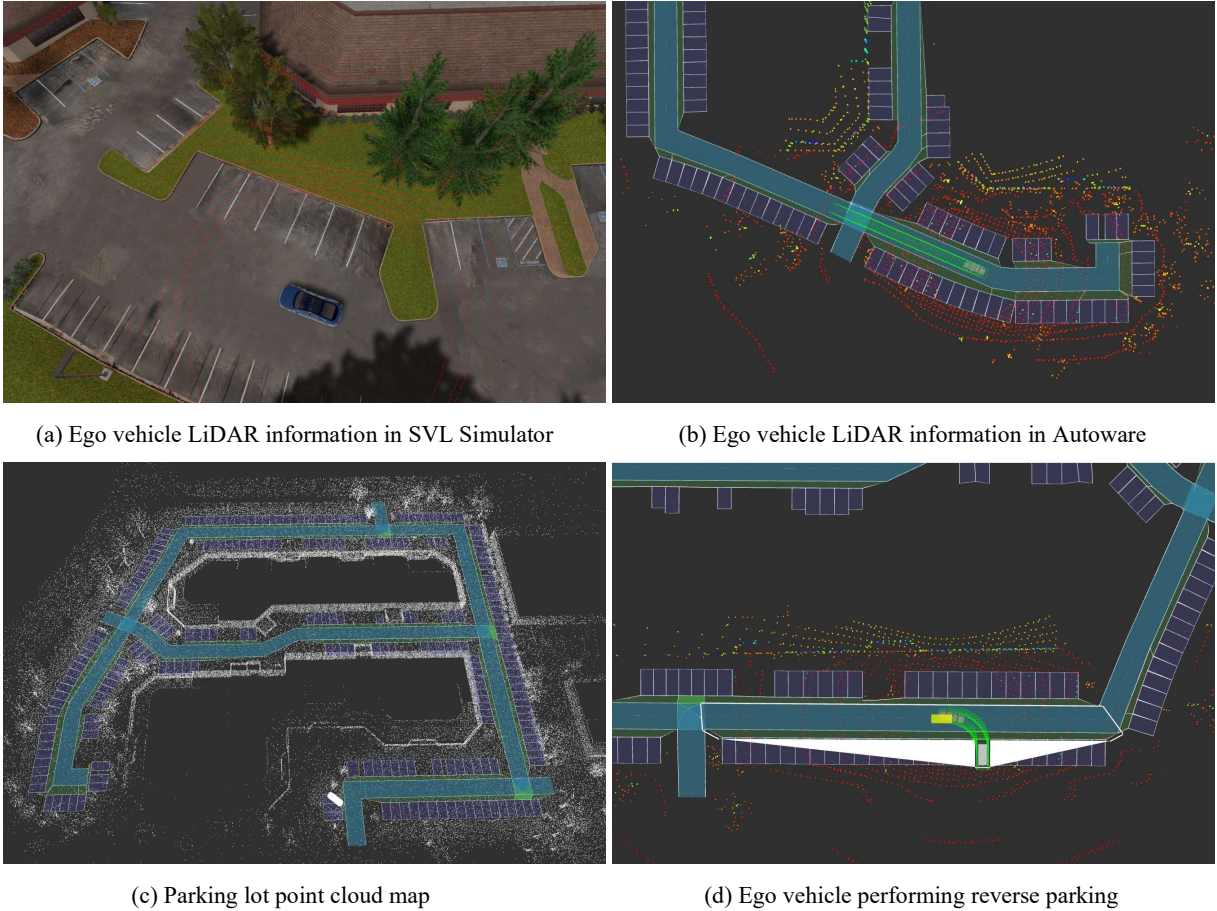

(a) Ego vehicle LiDAR information in SVL Simulator    (b) Ego vehicle LiDAR information in Autoware

(c) Parking lot point cloud map    (d) Ego vehicle performing reverse parking

**Figure 3.** Autoware and SVL Simulator synchronization demonstration

analysis of billions of miles on the road. We only attempt to show that AVs are capable of handling complex traffic conditions that can be challenging for human drivers.

To the best of knowledge, this is the first attempt to demonstrate the industrially employed AD platforms for educational purposes. We used both Apollo and Autoware because their interfaces with the SVL Simulator were good at demonstrating different features of an autonomous system. While both platforms supported advanced sensors such as the Camera, LiDAR, GPS, and IMU, Apollo's Dreamview interface displays the detailed object information captured by the camera (Figure 2), meanwhile Autoware's rviz2 interface shows a comprehensive image of the vehicle's LiDAR datapoints (Figure 3).

Another noticeable thing is that both Apollo and Autoware contain integrated hardware and software support, which means the same system can be run either on a real vehicle or in the simulation. The dual nature implies that the performance of the vehicle in the simulation has implications for the real world. This provides support for our study, as the participants could reflect on their simulator experience and gain insight on how AVs perform in real traffic. This connection has been further validated by the work of Fremont et al. In their research, various AV testing scenarios were set up at the GoMentum Station, and their digital twins were built and imported into the LGSVL Simulator (previous version of the SVL Simulator). AD was implemented using Apollo 3.5 and the researchers found that: "62.5 % of unsafe simulated test cases resulted in unsafe behavior on the track… 93.3 % of safe simulated test cases resulted in safe behavior on the track" [27]. While the simulation result did not exactly match real-world testing, this experiment clearly indicates that simulations can be mapped to the real world.

### IV. Experiment

*Survey*: we created a two-part survey to evaluate the effectiveness of our simulator. The first part was to be completed before the simulator experiment, and the second part afterwards. At the beginning of the first part, participants were asked to assess their understanding of AVs. For this question, we provided the following choices: "I hear about it from the news and social media", "I know the vehicle uses sensors and artificial intelligence but have no understanding of the technology", "I have some understanding of the different type of data collected by the sensors on an autonomous vehicle", "I have some understanding of the software and algorithms running on an autonomous vehicle". Then, participants were asked to answer twelve quantitative questions that measured perception of AVs from six perspectives: perceived risk, perceived usefulness, perceived ease-of-use, technical competence, situational management, and behavioral intention (Table 1). The questions were adapted from the AV acceptance model [5, 28] and used the seven-point Likert scale.

The second part was to be completed after the simulator demonstration. This part repeated the twelve quantitative questions asked earlier. This second survey allowed us to assess the evolution of opinion after the simulator experiment. Further, participants were asked to evaluate the usefulness (using the seven-point scale) of the driving information that was being displayed to them during the demonstration. The information included "the AV's planned routing on the map", "the AV's planning and control graph", "the AV's sensing of vehicles and pedestrians.", "the AV's prediction of vehicles and pedestrians", and "the AV's sensing of traffic indicators". The rest of the survey was mainly qualitative, focusing on the participants' opinions on the use of a driving simulator, their satisfaction with the AV's performance, and their remaining questions with autonomous vehicles.

*Procedure:* at the beginning of the study, participants were given a short introduction to the SVL Simulator. Then, they practiced driving through it using the steering wheel and pedal set. After they felt comfortable with the controls, participants were asked to drive through five pre-defined scenarios. For each scenario, we briefly introduced the map, environmental settings, and the route they needed to follow. Participants were not informed about the traffic emergency that would happen. Instead, they were told to pay attention to the road and drive safely. Each testing scenario took about two minutes, and the participants' driving performance was recorded. After the participants completed each scenario task, we demonstrated how AVs handled the same situation.

For the Apollo demonstration, we first mentioned that it was an industrially used AD platform and introduced the real-world Apollo projects. Then, we explained to the participants that our system is the same one running on a real vehicle, and the AV's performance in the simulator reflects what could happen in the real world. In the Apollo Dreamview interface, participants could watch the camera-detected vehicles, pedestrians, and their predicted movements. The main map showed the AV's routing information, and the panel on the right displayed the AV's planning and control graph (Figure 2). For each testing scenario, we also explained the AV's movements. For example, the AV stopped at the red light; the AV was yielding to pedestrians; the AV was switching lane because the current lane was blocked; the AV received a new destination and planned its route. The valet parking scenario was demonstrated using Autoware. Similarly, to Apollo, we first explained that this was another industrially used AD platform and showed the Autoware Lexus vehicle. During the simulator demonstration, participants watched the AV navigate from the drop-off point to the designated parking spot and perform reverse parking. Two important technological concepts were introduced: the LiDAR sensor and the High Definition (HD) Map. In the Autoware rviz2 interface, participants could observe the LiDAR image moving with the vehicle and reflecting the shape of the surrounding objects. We explained that this 360°, high precision laser sensor could detect obstacles and help avoid collisions. For the HD Map, we showed the parking lot with detailed road features and LiDAR-generated point clouds. The point clouds included houses, trees, and roadside shrubs. Participants were informed that this high-resolution map could provide valuable topographical information to the AVs and significantly boost their performance (Figure 3). The explanations were given in plain, non-technical language and, when combined with the simulator demonstration, were interesting and easily understandable to individuals with no technical background.

**Table 1:** Twelve quantitative questions

| | |
|---|---|
| Perceived Risk | I am worried about the safety of autonomous driving technology. |
| | I am concerned that failure or malfunction of the autonomous vehicle may cause accidents. |
| Perceived Usefulness | Using autonomous vehicles will increase my productivity. |
| | Using autonomous vehicles will increase my driving performance. |
| Perceived Ease-of-use | Learning to operate an autonomous vehicle would be easy for me. |
| | Interacting with autonomous vehicles would not require a lot of my mental effort. |
| Technical Competence | I believe that autonomous vehicles act consistently, and their behavior can be forecast. |
| | I believe that I can form a mental model and predict future behavior of an autonomous vehicle. |
| Situational Management | I believe that autonomous vehicles are free of error. |
| | I believe that autonomous vehicles will perform consistently under a variety of circumstances. |
| Behavioral Intention | I intend to ride in autonomous vehicles in the future. |
| | I expect to purchase an autonomous vehicle in the future. |

## V. Results

*Awareness of Autonomous Driving:* of the 28 participants, 7.1 % answered that they have heard about AVs in the news and social media, 42.9 % stated that they knew the vehicle used sensors and artificial intelligence, but had no understanding of the technology, 39.3 % stated that they knew about the different types of data collected by the sensors on an autonomous vehicle, and 10.7 % said they had some understanding of the software and algorithms running on an autonomous vehicle. This result was desired, as our demonstration was designed for people with limited understanding of the AD technology.

*Driving Performance:* in the vehicle following scenario, 42.9 % of the participants collided with the vehicle in the front, and 32.1 % had at least one near-collision case. In comparison, the Apollo autonomous vehicle maintained a safe distance throughout the way and responded almost simultaneously to the front vehicle's speed changes. In the lane block scenario, 21.4 % of the participants collided with the illegally stopped vehicle, and 17.9 % collided with the passing vehicles on the left when attempting to switch lanes. In this test, Apollo detected the illegally stopped vehicle from afar and switched lanes after ensuring safety. In the pedestrian jaywalking scenario, 46.4 % of the participants failed to stop the vehicle and hit the pedestrian. In contrast, Apollo activated an emergency brake as soon as the pedestrians appeared and avoided the accident. In the city traffic scenario, all of the participants reached the destination safely. In this scenario, the Apollo vehicle stopped at the unprotected left turn, yielded to other agents, and then proceeded with caution. The valet parking scenario was not intense and successfully completed by all participants. This task was also achieved using Autoware.

*Quantitative Questions:* participants' responses to the twelve quantitative questions before and after the simulator experiment were categorized into six measurements and analyzed using the Paired-

**Table 2**. Quantitative measurements before and after the simulator experiment

| Measure | Before simulator | | After simulator | |
|---|---|---|---|---|
| | M | SD | M | SD |
| Perceived risk | 5.214 | 1.139 | 3.857 | 1.406 |
| Perceived usefulness | 5.143 | 1.307 | 5.714 | 1.437 |
| Perceived ease of use | 5.357 | 0.969 | 5.571 | 1.254 |
| Technical competence | 5.143 | 1.216 | 5.393 | 1.347 |
| Situational management | 3.714 | 1.032 | 4.393 | 1.521 |
| Behavioral intention | 5.571 | 1.222 | 5.536 | 1.365 |

Samples T Test. After the simulator demonstration, we observed a significant decrease $t(26) = 2.994, p = 0.011$, in the participants' perceived risk of autonomous vehicles. Meanwhile, there was a significant increase in the perceived usefulness of the AD technology, $t(26) = -2.327, p = 0.040$. We also observed some increase in the AV's score of situational management, $t(26) = -1.494, p = 0.161$, and technical competence, $t(26) = -1.298, p = 0.219$, while the result was not statistically significant. Participants' views on the AV's perceived ease-of-use did not change by much, $t(26) = 0.424, p = 0.679$, and their intention to use an autonomous vehicle remained almost the same, $t(26) = -0.200, p = 0.844$ (Table 2).

*Information Evaluation:* participants also rated the usefulness of the AD information that was being displayed during the demonstration. Of the five choices, the AV's routing information ($M = 6.071, SD = 0.997$), sensing of vehicles and pedestrians ($M = 6.143, SD = 1.292$), sensing of traffic indicators ($M = 6.071, SD = 1.141$), and prediction of the vehicles and pedestrians ($M = 6.214, SD = 0.802$), all received an average score above six. These were very high ratings, as the maximum score was seven. The AV's planning and control information received a slightly lower score ($M = 5.214, SD = 2.007$), which was justified as this part involved some technical terms that require professional knowledge for understanding, such as the "Planning Theta", "V-T Graph", and "Kappa Derivative".

*Qualitative Questions:* in the qualitative questions, most participants stated they were very satisfied with the AV's performance. They believed the AV was equipped with advanced software and hardware that were capable of handling complicated situations. Some answered that this simulator experience provided them with great insight of how AVs work, while others mentioned this simulated experience encouraged them to ride in an actual vehicle when provided with the opportunity. There were several questions and concerns as well. Some participants complained that the AV drove too cautiously and proactively, and that it may not be the most efficient way of transportation. Some stated that because the AV put considerable emphasis on safety, human drivers may deliberately take advantage of them by cutting them in lane or not yielding. Despite all the disputes, all our participants seemed to agree that the AV had a high score of safety and provided an alternative way of transportation.

## VI. Discussion

*Driving Performance*: while our autonomous control produced significantly better driving results, a lot of factors could be affecting the performance of human drivers. First, the driving experiment was done in third person view, which was unconventional to a human driver. Also, our stationary workstation could not generate the movements and forces one would normally experience in a real car, and therefore may not be immersive enough.

While the limitations exist, note that the primary purpose of this experiment was not to show that AVs drive better than humans, but to help human drivers build a correct perception of the AVs. After driving through the testing scenarios and watching how AVs handled the same situations, the participants could get the sense that: the AVs can react quickly to an emergency, navigate safely along a pre-defined route, and perform valet parking in a learned space.

*Survey Feedback*: we were not surprised to find that our simulator demonstration reduced the participants' perceived risk of autonomous vehicles. In all of our testing scenarios, the AV acted very cautiously and responded to emergencies without hesitation. Further, the vehicle's routing, planning, sensing, and control information was quite self-explanatory and clear enough to reassure the participants. This was demonstrated in the participants' high ratings of the AD information.

While a detailed demonstration of the AD technology reduced the participants' perceived risk, many found that it required a lot of their mental effort to keep track of the system information and be cognitive of the vehicle's state. As a result, the participants' ratings on the AV's perceived ease of use did not increase. Some drivers initially assumed that interacting with an autonomous vehicle was going to be easy. However, they soon found there was still information that required their attention or even intervention. Despite the higher cognitive load, drivers still preferred for the information to be displayed, as it increased the system transparency and helped them understand the vehicle's movements.

We also observed different levels of increase in the vehicle's perceived usefulness, technical competence, and situational management. Since the AV handled all five testing scenarios smoothly when compared to human drivers, many participants believed that the AV would increase their driving performance and productivity. In addition, some participants felt that they could form a mental model to understand and predict the vehicle's behavior. With a more in-depth understanding of the technology, participants showed less concern towards system malfunction and believed in its consistency and reliability under a variety of circumstances.

Despite a significant decrease in the perceived risk and a significant increase in the perceived usefulness, we did not observe an increase in the participants' intention to use an autonomous vehicle. This finding differed from our initial expectations, and we attribute it to two reasons. First, the participants' intention to use an autonomous vehicle was already high before our simulator demonstration. While our participants had limited understanding of AD technology, they believed that the technological shift is inevitable and expected to use an autonomous vehicle in the future. Second, there is a lack of commercially used AVs on the market (SAE Level 3 to 5), and an absence of laws and government regulations. While our participants embraced the idea of AVs, they did not feel like there was an imminent need to use or purchase one.

*AD Failure Cases*: a final point to mention is that some AD failures, although not life threating, were not covered in our demonstration. During our engineering test and scenario development, we found that the AV consistently failed to handle some challenging cases such as: traversing the center line to avoid a stopped vehicle, deliberately cutting in front of other vehicles to make a turn, or navigating from some private, unmarked area to the main road. While we were aware of the limitations of the AV, they were not presented to participants as the primary purpose of this experiment was education and trust development. Meanwhile, we believe these technological imperfections should be recorded and reported to the industrial developers so they can fix the problems and improve the next generation of AVs. In the future, we will work on identifying the regular and edge cases and focus on human-AV interaction to help drivers further understand the merits and limitations of the technology.

## VII. Conclusion

We presented in this study a driving simulator and testing scenarios to improve people's understanding and trust in autonomous vehicles. To that effect, we leveraged the industrially used autonomous driving platform Apollo and Autoware. Our study of 28 participants showed that this system successfully reduced the perceived risk and increased the perceived usefulness of autonomous vehicles. There were limitations as well. First, our five testing scenarios are not representative of all traffic conditions and require further development. Also, our third person view simulation could be changed to first person view, and implemented in more immersive environments such as virtual reality or augmented reality. Furthermore, our small number of participants should be regarded as a pilot towards the development of a larger study, where perception can be correlated with age and gender. Overall, our simulation system acts as a low-cost and reliable platform for autonomous driving testing and demonstration.